\documentclass[aps,pre,twocolumn,a4paper,floatfix,showpacs]{revtex4}
\usepackage{graphicx}
\usepackage{dcolumn}
\usepackage{bm}
\usepackage{hyperref}
\usepackage{rotate}
\usepackage{color}
\newcommand {\eps} \epsilon
\newcommand {\comas} [1] {\textquoteleft #1 \textquoteright}

 \usepackage{epstopdf}
\usepackage{epsfig}

\begin{document}

\title{Tuning synchronization of integrate-and-fire oscillators through mobility}

\author{L. Prignano}
\author{O. Sagarra}
\author{A.  D\'{\i}az-Guilera}
\affiliation{Departament de F\'{\i}sica Fonamental, Universitat de Barcelona, 08028 Barcelona, Spain}
\begin{abstract}
We analyze the emergence of synchronization in a population of moving integrate-and-fire oscillators. Oscillators, while moving on a plane, interact with their nearest neighbor upon firing time. 
We discover a non-monotonic dependence of the synchronization time on the velocity of the agents. Moreover, we find that mechanisms that drive synchronization are different for different dynamical regimes. We  report the extreme situation where an interplay between the time-scales involved in the dynamical processes completely inhibit the achievement of a coherent state. We also provide estimators for the transitions between the different regimes.
\end{abstract}
\maketitle


After more than one decade of research on complex networks\cite{ab02,blmch06}, where they have been understood as simple collections of nodes and   (weighted) links\cite{bbpv04}, time has come to analyze new paradigms of complex topologies. Examples of these new settings are, for instance, time dependent networks\cite{hs12}, networks at different layers (usually called multiplex networks) \cite{mrmpo10}, spatial networks \cite{barth11} and interdependent networks \cite{bppsh10}. 

Concerning the case of time-dependent networks, up to now the interest has been focused on the need of finding new tools for characterizing topological properties of networks whose connectivity changes in time, but not as much on the interrelation between topological changes and dynamical evolution of the nodes themselves. 
This interrelation has been mainly considered for the asymptotic cases in  which characteristic times of both dynamical processes are very different. When topological changes are very fast, a good proxy is the fast switching approximation (FSA) \cite{frasca08} whereas in the opposite limit, perturbation around the static case makes sense.
 In this paper we analyze dynamical effects in time dependent networks but understanding the network as the result of the motion of agents that interact when they are close enough\cite{bffr06,tjp03,bschdms06}. 
 
When considering emergent properties of systems formed by elementary units with its own dynamics, one of the most important ones is synchronization \cite{prk01}. Actually, synchronized behaviors appear in nature in groups that improve performance based on collaboration \cite{strogatz03} and also in human actions it has shown to provide collective benefits \cite{hp04,shu11}.
The effect of changing patterns of interaction on synchronization features has been analyzed in different settings, for instance in chemotaxis \cite{t07}, mobile ad hoc networks \cite{r01}, wireless sensor
networks \cite{sy04} and the expression of segmentation clock genes
\cite{umi10}.

Recently, a general framework of mobile oscillator networks
where agents perform random walks in a two-dimensional (2D) plane has been proposed \cite{fkd11a}.
This framework, that reduces to FSA when velocity is high enough, is valid for models whose evolution can be well approximated by linear dynamics. This actually holds for models such as populations of Kuramoto oscillators \cite{kuramoto84,abprs05},  whose evolution, after a short transient time, is very well described by a set of linear equations that can be solved in terms of spectral properties of the Laplacian matrix \cite{fkd11b}. Alternative approaches based on Fokker-Planck equations have been also proposed recently \cite{pnm10}.

Here, we focus on a dynamical system, a population of integrate and fire oscillators (IFO), where the evolution takes place in two different time scales. One for the slow evolution of the internal state variables (the phase and the orientation) and the other for the instantaneous interaction between the units (pulse coupling). 
During the last years it has been shown that the  interaction structure plays a fundamental role in the dynamics of IFO networks \cite{adkmz08}.
Usually, IFO have been used to model neural systems but we can also find some examples of applications in other fields, as for example in economy \cite{edga11}. 

In the present letter we consider a population of integrate-and-fire oscillators, which interact with their nearest neighbor only while freely displacing on a plane. 
Such a minimal interaction rule, in contrast to other approaches based on an interaction radius \cite{psgd12}, provides a strong non-monotonic dependence of the synchronization time on the velocity of the agents.
Three different regimes are identified according to the synchronization properties of the system: a slow regime, a fast switching limit and an anomalous intermediate region between them.
Remarkably, in this last region a divergence of the synchronization time is observed.
Depending on the studied phenomena, synchronization can be taken either as as a positive or negative feature of the collective behavior of the system. Global brain synchronization, for instance, is associated to epileptic seizures whereas local synchronization is related to some cognitive tasks \cite{mk12}. Synchronization in trading activities can be harmful for the stability of financial systems \cite{shu11}. Thus designing mechanisms that prevent global synchronization can become a very important tool in complex dynamical systems \cite{laah12}.

\begin{figure}[th]
\begin{center}
\includegraphics[width=\columnwidth]{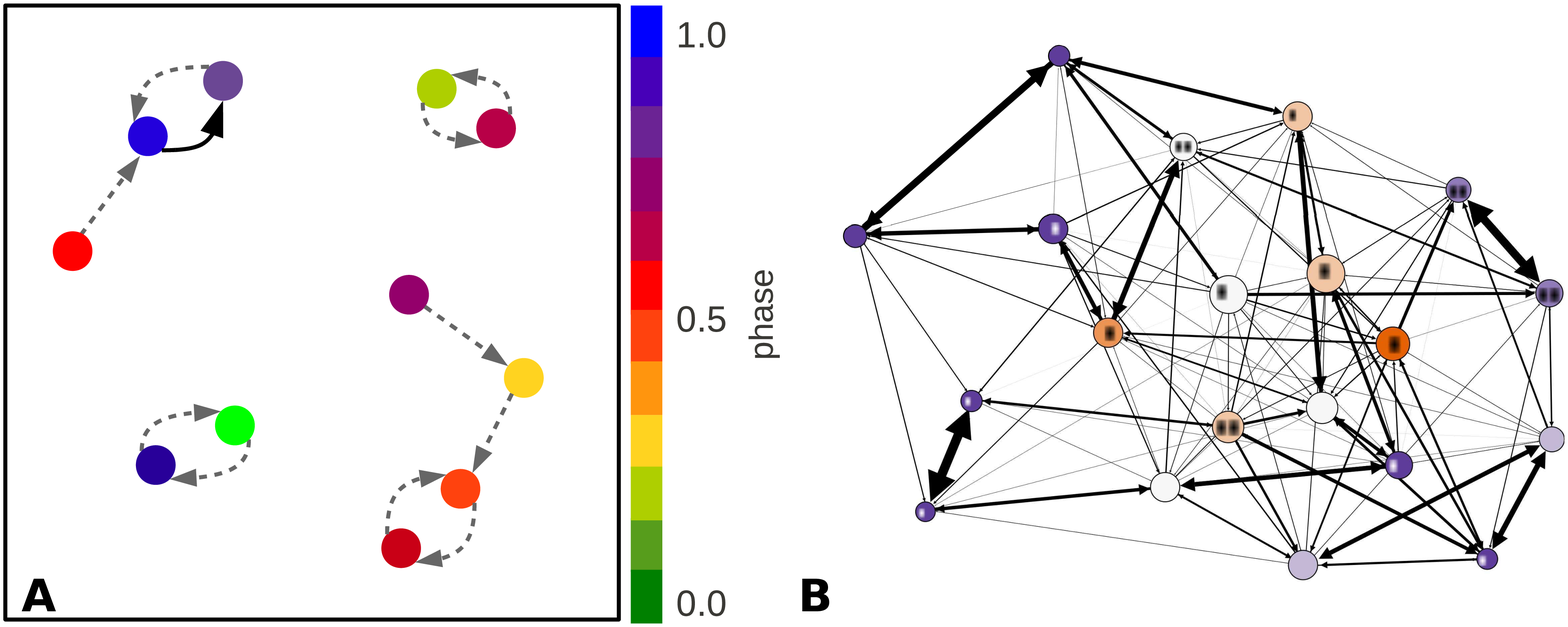}
\includegraphics[width=\columnwidth]{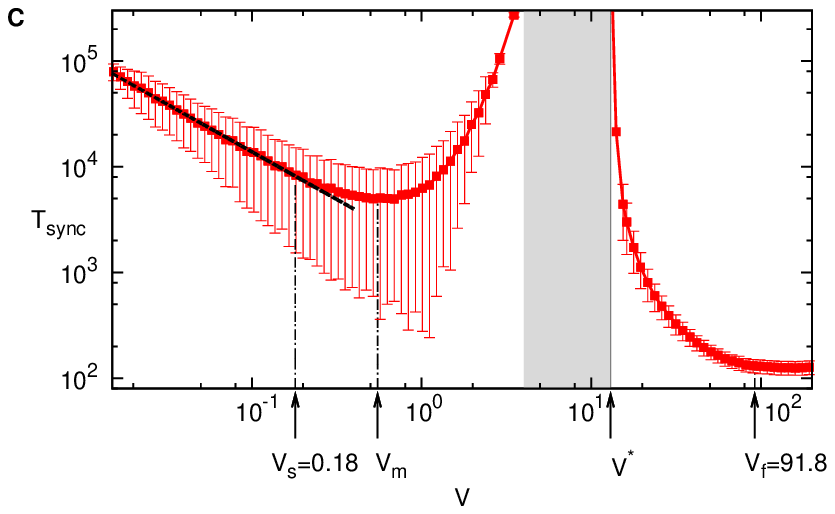}
\caption{\textbf{A} Snapshot of the system in the incoherent state. Each disk is an oscillator, gray dashed arrows stand for first-neighbor relationships, while the continuous black one represents a firing event that is taking place at that precise instant. \textbf{B} Final ($T=T_{sync}$) total interaction network at $V=V_m$. Each node is an oscillator. Links represent shots, their weights are proportional to occurrence of the interactions among neighbors. Node color changes from purple to orange increasing the in-degree. Size increases with increasing out-degree.
\textbf{C} The average synchronization time $T_{sync}$ as a function of $V$, for $L=400$, $N=20$, $\eps=0.1$. Values of $V_f$ and $V_s$ are calculated using estimators (\ref{vf}) and (\ref{vs}) respectively. In the following, when not otherwise states, the values of the parameters are those used in this figure. Averages are performed over $2000$ realizations and errorbars correspond to one standard derivation.}
\label{fig0}
\end{center}
\end{figure}

Our model consists on a population of $N$ moving oscillators with identical velocity $V$ and random orientation on a square of side length $L$ with periodic boundary conditions. The internal phases of the agents $\phi \in (0,1)$ increase uniformly with period $\tau$,
\begin{equation}
\frac{d\phi_i}{dt}=\frac{1}{\tau},
\end{equation}
until they reach a maximum value of 1, when a firing event occurs and the phase is reset. Upon such an event at time $t$, the firing oscillator influences its \emph{nearest neighbor} (oscillator at minimal distance, labelled nn, see Fig. \ref{fig0}A) producing a random reorientation of its motion and an update in its phase by a factor $\eps$:
\begin{equation}
\phi_{i}(t^-)=1 \Rightarrow \left \{ \begin{array}{l} \phi_i(t^+)=0\\ \phi^i_{\text{nn}}(t^+)=(1+\eps)\phi^i_{\text{nn}}(t^-) \\ \theta^i_{\text{nn}}(t^+) \in  [0,2\pi]\end{array} \right. ,
\end{equation}

The phase update is performed at frozen time until the phases of all oscillators have been updated (some agents may reach their threshold and fire upon receiving a phase update from a firing neighbor). Then the phases evolve again uniformly in time until another update is triggered.

The system is synchronized when a succession of consecutive firing events (avalanche) equal to the system size $N$ is detected. For the sake of clarity we define the (discrete) time $T$, as the number of times a given oscillator (that we identify with oscillator 0 in our computer simulations) has fired. This allows us to define $T_{sync}$ as the number of cycles this reference oscillator takes to enter the synchronized state (i.e., the number of updates needed for an avalanche of size $N$ to occur).


The chosen minimal interaction rule is such that the system lies far below the static percolation transition \cite{dc02,bbw05}. Therefore, global synchronization is not achievable without motion (since it is very unlikely that a giant connected component of size $N$ exists \footnote{A giant connected component is just a necessary condition, but generally a static configuration does not allow the system to synchronize.}). 
It is the non-null velocity of the oscillators that enables the system to reach the coherent state, therefore we could expect $T_{sync}$ (the average time the system needs to synchronize) to be a decreasing function of $V$, such that $T_{sync}\to \infty$ when $V \to 0$ and $T_{sync}\to T_f>0$ (a constant value) when $V$ is high enough \footnote{When $V$ is such that the agents, at each time step, cover a distance of the same order of magnitude as of the linear dimensions of the box, then the interactions are completely randomized (fast switching approximation). A further increasing of $V$ does not produce any effect.}.

Fig.\,\ref{fig0}C shows that, for $V<V_s$, the synchronization time decreases as a power of $V$ when the latter increases. Then, the decreasing slows down and $T_{sync}$ has a minimum at $V=V_m>V_s$. Beyond this value, the synchronization time gets larger and larger, until the system enters a region where it is unable to reach the coherent state in a finite time (gray area in Fig.\,\ref{fig0}C). For even larger values of the velocity, $V>V^*$, it decreases abruptly, finally reaching its asymptotic value when $V=V_f$. 

Three main regions are thus identified: A \comas{no-synchronization zone} in the middle that separates a \comas{left region} (small $V$) and a \comas{right region} (high $V$). In the left region, we can separate two sub-regions: on the left, at $V<V_s$, there is what we call the \emph{slow regime}, and, on the right, a transition zone. The same happens for the right region. On the left, we find a transition region, while on the right, at $V>V_f$, the system enters the \emph{fast limit}, where $T_{sync}$ no longer depends on $V$.


Before entering into a more specific discussion we present a general characterization of the system behavior. 

\begin{figure}[ht]
\begin{center}
\includegraphics[width=\columnwidth]{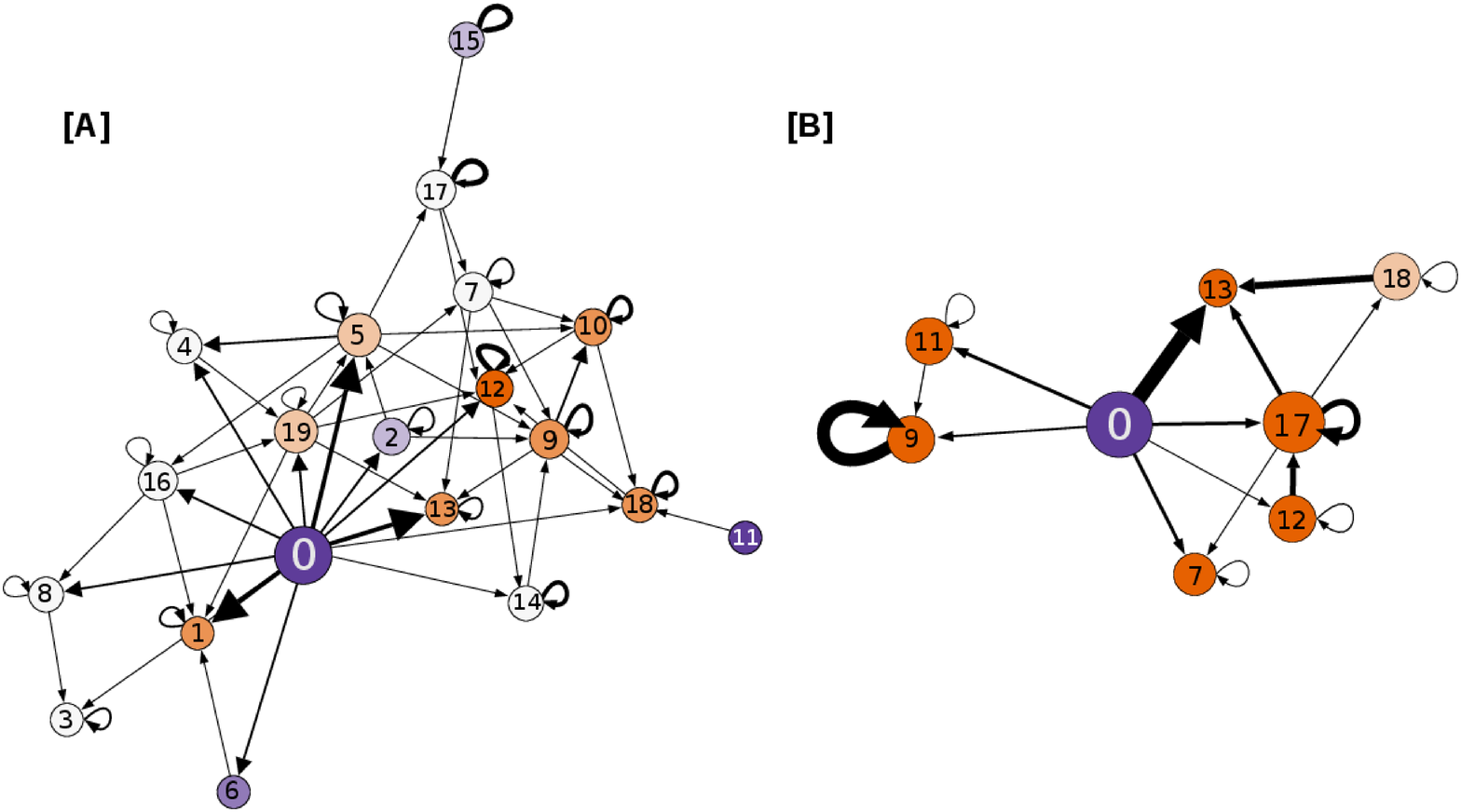}
\includegraphics[width=\columnwidth]{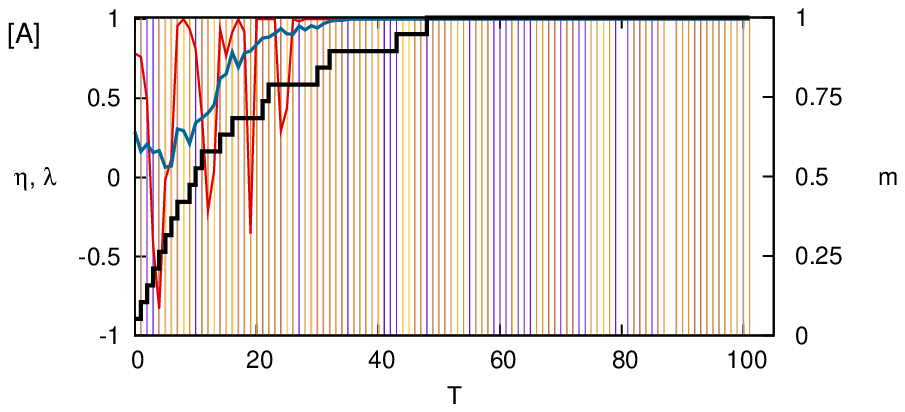}
\includegraphics[width=\columnwidth]{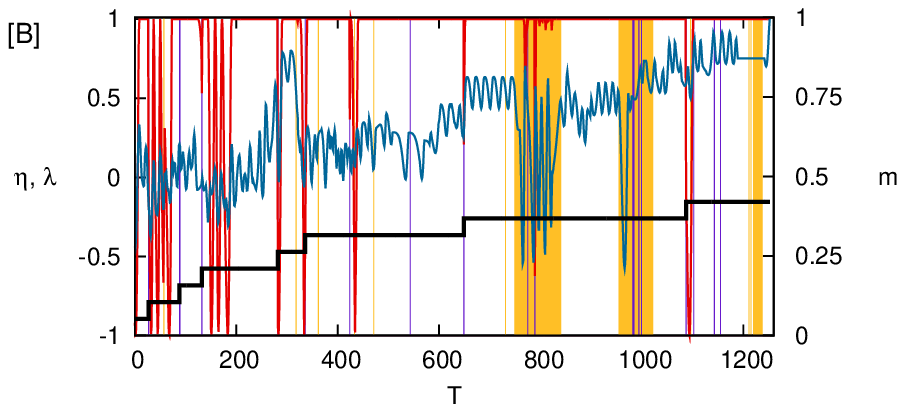}
\caption{\textbf{Upper panel:} Final ($T=T_{sync}$) network of the interactions mediated by a single oscillator (labelled "0"), in the fast limit, at $V=2V_f$ (panel A) and at $V=V_m$ (panel B) respectively. Node and link color and size code as in figure 1 B.
\textbf{Lower panels:} 
Parameters $\eta$ (blue), $\lambda$ (red) and $m$ (black) are plotted as a function of time, from $T=0$ to the synchronization time for a given oscillator and a single realization of the process. Yellow vertical lines mark a change of the in-neighbor, while the purple lines stand for a change of the out-neighbors. In panel A, the velocity is $V=2V_f$. In  panel B, it is $V=V_m$. Where yellow lines are so dense that they form a yellow band, it means that the oscillator has more than one in-neighbor simultaneously.}
\label{fig2}
\end{center}
\end{figure}

In Fig.\,\ref{fig2}, upper panels, we introduce the cumulative individual interaction network (CIN) of an oscillator (labelled $0$), for two independent synchronization process at two different velocities (panel C: $V=2V_f$ while panel D: $V=V_m$ ).
A CIN represents in a visual way the role played by a given unit in the \emph{signal spreading} and is constructed in the following way: whenever an oscillator $i$ fires at oscillator $0$ and oscillator $0$ fires at oscillator $j$, a link between $i$ and $j$ is added. If the link already exists, its weight is increased. In the case $0$ does not receive any shot, we put a link between $0$ and $j$. A reciprocal shot is represented as a self-link. We repeat this process until the system reaches the synchronized state.

To better understand the synchronization mechanisms we also report different relevant magnitudes.
In Fig.\,\ref{fig2}, lower panels (panel A: $V=2V_f$, panel B: $V=V_m$ ) we plot
the global order parameter $\eta(T)=\langle\cos{(2\pi\phi_i(T))}\rangle$ and in order to have an insight about what happens at local scale, the quantity $\lambda(T)=\cos{(2\pi\phi^0_{\text{nn}}(T))}$ 
where $\phi^0_{\text{nn}}$ 
is the phase of the unit to which the reference oscillator fired at time $T$. These parameters measure the synchrony of our system, ranging from a uniform phase distribution of our oscillators ($\eta(T)\approx\lambda(T) \approx 0$) to complete synchronization ($\eta(T)=\lambda(T)=1$). Note that the averages are calculated with respect to all oscillators' phases $\{\phi_i\}$ in the case of the global order parameter and only with respect to the nearest neighbor's phase  $\phi^0_{nn}$ in the case of the local one \footnote{Note that the average is calculated upon a firing event by the reference oscillator, hence our order parameters are an average of the phase difference of the other oscillators (only nearest neighbors in the case of $\lambda(T)$) with respect to this one.}. 

Finally, $m(T)$
represents the total fraction of oscillators that have been out-neighbors of the reference oscillator up to time $T$.

We observe strong differences between the two scenarios.
When agents move fast, all nodes appear represented in the CIN since they play a global role sending and receiving signals throughout the whole system. Interactions are completely rewired at each time $T$, therefore $m(T)$ increases very rapidly and $\phi_{nn}$ is just a random variable extracted among $N-1$ possible ones. This means that $\lambda$ is exactly the same as $\eta$, but with less statistics. Both quantities increase together (more or less noisily) because by means of the firing events the whole phase distribution becomes narrower.

In contrast, when the mobility of the agents is reduced, few nodes conform the CIN. Each oscillator plays a local role mediating the interactions among a small number of units that are the same all the time, no matter how long the synchronization process could be. The behaviors of $\eta$ and $\lambda$ appear to be uncorrelated, being $\lambda=1$ almost all the time: each oscillator spends a long time with its neighbors, usually being able to synchronize with it before changing. At the beginning ($T<500$) whenever it starts firing towards a new oscillator (black vertical lines in Fig.\,\ref{fig2}) $\lambda$ experiences an abrupt decreasing while $m(T)$ increases as it is the first time the reference oscillator \emph{meets} that neighbor. Later ($T>500$), the chances to change a neighbor for another one already known, having a very similar phase, increase. The phase distribution becomes narrower and specially at local scale, among the units the oscillator of reference can meet, the dispersion is small. Consequently neighbor changes do not affect $\lambda$ anymore.

The observations point out that  the fast regime can be understood as an \emph{homogeneous} regime while slow velocity enhance heterogeneity among units. Hence the mechanisms that allow the system to synchronize have to be different in the two cases. The system has different \emph{strategies} to reach the coherent state in the left and in the right region but neither of them work in the intermediate region.

In the following paragraphs we will determine the region of the parameters space corresponding to each regime providing a quantitative estimator for both $V_s$ and $V_f$ as functions of $N$, $\eps$, and $L$. 

In order to understand the different regimes we need to compare the time scales of the dynamical processes that lead the system to its final synchronized state, i.e. how fast are oscillators to change neighbors effectively and how fast they are able to locally synchronize.

Let us first focus on the fast regime. As a kind of mean-field assumption we can imagine a single oscillator, $i$, moving on a plane where the rest of the population is fixed in the most disperse possible configuration, a square lattice such that the
system is divided into $(N-1)$ squares of side length $\ell =L/\sqrt{N-1}$. Oscillator $i$ changes its neighbor when it exits a square to enter one of the adjacent ones.
It will do so between time $T-1$ and $T$ if, at time $T-1$, the component of $V$ perpendicular to that side is larger than the distance s separating $i$ from that same boundary.
First, notice that if $V\gtrsim \ell$ then the exit probability is equal to 1 and each unit changes its neighbor at each time step. One can conclude that $\ell$ is a good estimator for the value of the velocity at which the system enters the fast switching limit, i.e. 
\begin{equation}
V_f=L/\sqrt{N-1}.
\label{vf}
\end{equation}

On the opposite limit, when oscillators move very slowly ($V\sqrt{N}/L\ll1$), we can also compute the average time spent in changing neighbors (see the Supplemental Material \cite{suppmat})
\begin{equation}\label{t-out}
T_{out}=\frac{\pi L}{4V\sqrt{N-1}}.
\end{equation}

This time needs to be compared with the time a characteristic cluster needs to locally synchronize \cite{fkd11a}. Given the minimal rule of interaction we have proposed, the average size of (weakly) connected components in the resulting network of our model is  $S_{cc}=(8\pi+3\sqrt{3})/3\pi\approx 3$ (see \cite{ps85} and \cite{suppmat}). For such a characteristic cluster of size 3 there is only one possible configuration\,\footnote{The only other 3 oscillators configuration has probability zero since all distances among pairs of oscillators need to be equal.} (see Fig. \ref{fig0}A) whose synchronization time is given empirically by the expression 
\begin{equation}\label{t3}
T_3=\kappa/\eps .
\end{equation}
with $\kappa=2.0\pm 0.1$ \cite{suppmat}.
A necessary condition for the system to reach complete local synchrony is that, on average, no topological change has to occur during a time frame $T_3$.
Since the average time separation between two neighbor changes of the same unit is $T_{out}$, the average time separation between two topological changes in the system can be expressed as $T_{out}^{(N)}=T_{out}/N\simeq \pi L/(4V N^{3/2})$. 

With these two ingredients, we can easily identify the starting point of the slow regime as that set of parameters values such that 
$T_3 \simeq T^{(N)}_{out}$.
Hence, fixing $N$, $\eps$ and $L$, we obtain
\begin{equation}\label{v_s}
V_s= \frac{\pi}{4\kappa}\frac{\eps L}{N^{3/2}}.
\label{vs}
\end{equation} 
 
\begin{figure}[ht]
\begin{center}
\includegraphics[width=.9\columnwidth]{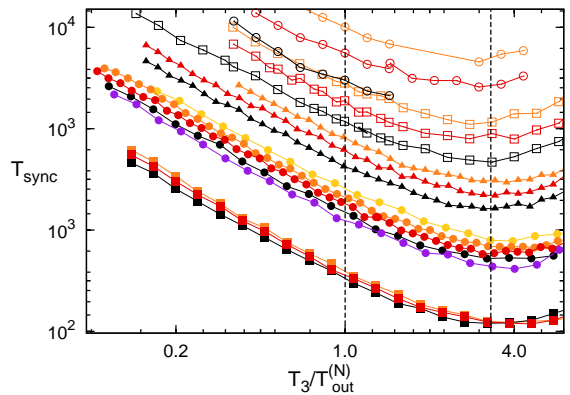}
\caption{Average $T_{sync}$ as a function of the ratio $T_3/T_{out}^{(N)}$ for several values of $N$ (filled squares: $N=10$, filled circles: $N=20$, triangles: $N=30$, empty squares: $N=40$, empty circles: $N=50$) and $\eps$. The considered values of $\eps$ are, from the top to the bottom: $\eps=0.05$ (orange), $\eps=0.1$ (red), $\eps=0.2$ (black); additionally, for $N=20$, we also considered $\eps=0.02$ (yellow top curve) and $\eps=0.3$ (purple bottom curve). The vertical dashed line at $T_3/T_{out}^{(N)}=1$ corresponds to $V=V_s$, the other one to $V=V_m$. Averages have been performed over $2000$ (for $N\leq20$) or $1000$ ($N>20$) realizations.}
\label{tslow}
\end{center}
\end{figure}

Both $V_f$ and $V_s$ are reported in Fig.\,\ref{fig0} and have been checked for various values of $\eps$, $N$, and $L$, showing a very good agreement. In particular, in  Fig.\,\ref{tslow} we show how our predictor mark the transition point of the slow regime. Moreover, we can identify a universal minimum for all the curves marking the value of $V_m$.

Summarizing, we have proposed a model of moving integrate-and-fire oscillators that interact only with the nearest neighbor. This minimal interaction rule can account for some physical situations where communication between agents is minimized. We have found three very different dynamical regimes. In the region where oscillators move very slowly and oscillators tend to keep the same neighbors for a long time, local synchronization dominates and a global one proceeds through a very large number of interactions (firings and changes of neighbors).
In the other limit, of very fast motion, oscillators fire at random neighbors which makes that interactions to be very effective and phases approach each other in a monotonous way and synchrony is achieved directly on the global scale.
Interestingly, we find an intermediate regime where none of these mechanisms work and the system is not able to reach synchronization. 
The presence of a no synchronization band needs to be further studied and its robustness checked for other types of non-linear interacting oscillators in future works.

Despite synchronization is usually seen as a positive outcome of a cooperative dynamical system, under some circumstances, as it happens, for instance, in brain dynamics, global synchronization is not a healthy state. 
Our model represents a paradigmatic example of a system where synchronization can be prevented by tuning the rate at which the topology changes. 
We have determined  the bounds of this region in terms of the characteristic times of the model, those related to local synchronization and to mobility.


\acknowledgments This work has been partially supported by the
Spanish DGICYT Grant FIS2009-13364-C02-01  and by the Generalitat de Catalunya 2009-SGR-00838. 
L.P. and O.S. have been supported by the Generalitat de Catalunya through the FI Program.

\bibliography{biblio}
\end{document}